\documentclass[a4paper,11pt]{article}
\usepackage{jheppub} % for details on the use of the package, please see the JINST-author-manual
\usepackage{lineno}
\usepackage{xspace}
\usepackage{comment}
\usepackage{subfigure}
\newcommand{\jpsi} {\ensuremath{{\mathrm J}/\psi}\xspace}
\newcommand{\psip} {\ensuremath{\psi'}\xspace}

\newcommand{\pp}           {pp\xspace}

\newcommand{\pA}           {\mbox{p--A}\xspace}
\newcommand{\AAcoll}           {\mbox{A--A}\xspace}

\newcommand{\nineH}        {$\sqrt{s}~=~0.9$~Te\kern-.1emV\xspace}
\newcommand{\seven}        {$\sqrt{s}~=~7$~Te\kern-.1emV\xspace}
\newcommand{\eight}        {$\sqrt{s}~=~8$~Te\kern-.1emV\xspace}
\newcommand{\twoH}         {$\sqrt{s}~=~0.2$~Te\kern-.1emV\xspace}
\newcommand{\twosevensix}  {$\sqrt{s}~=~2.76$~Te\kern-.1emV\xspace}
\newcommand{\five}         {$\sqrt{s}~=~5.02$~Te\kern-.1emV\xspace}
\newcommand{\fiveExactly}  {$\sqrt{s}~=~5$~Te\kern-.1emV\xspace}
\newcommand{\twosevensixnn}{$\sqrt{s_{\mathrm{NN}}}~=~2.76$~Te\kern-.1emV\xspace}
\newcommand{\fivenn}       {$\sqrt{s_{\mathrm{NN}}}~=~5.02$~Te\kern-.1emV\xspace}

\newcommand{\GeVc}         {Ge\kern-.1emV/$c$\xspace}
\newcommand{\MeVc}         {Me\kern-.1emV/$c$\xspace}
\newcommand{\TeV}          {Te\kern-.1emV\xspace}
\newcommand{\GeV}          {Ge\kern-.1emV\xspace}
\newcommand{\GeVtwo}       {Ge\kern-.1emV$^2$\xspace}
\newcommand{\MeV}          {Me\kern-.1emV\xspace}
\newcommand{\GeVmass}      {Ge\kern-.2emV/$c^2$\xspace}
\newcommand{\MeVmass}      {Me\kern-.2emV/$c^2$\xspace}

\title{\boldmath Zero-bias new particle searches using autoencoders in UPCs and diffractive events}

\author[1]{S. Ragoni\note{Corresponding author.}}
\author[]{, J. Seger}
\author[]{, C. Anson}
\affiliation{Creighton University,\\2500 California Plz, Omaha, \\NE 68178, United States, USA}

\emailAdd{simone.ragoni@cern.ch}
\emailAdd{jseger@creighton.edu}
\emailAdd{chrisanson5@gmail.com}

\abstract{We present an application of unsupervised learning for zero-bias detection of rare particle decays and exotic hadrons in low-background environments such as those characteristic of diffractive events and ultraperipheral \pp, \pA, or \AAcoll collisions at the CERN Large Hadron Collider (LHC), or in e--A collisions at the ePIC experiment at the future Electron-Ion Collider (EIC). Using a toy dataset simulating the decays of known resonances, including \jpsi and \psip, as well as more exotic candidates, we implement an autoencoder neural network to identify anomalies in the decay kinematics. The autoencoder, trained solely on typical events, is designed to reconstruct normal decays with low error while flagging anomalous decays based on the reconstruction error. We demonstrate that the autoencoder successfully separates typical decays from rare exotic events, with peaks in the invariant mass distribution corresponding to the injected rare signals. Our method shows promise in detecting rare, unpredicted processes in large-scale collider data, offering an effective approach for discovering new physics beyond the Standard Model.}

\begin{document}
\maketitle
\flushbottom

\section{Introduction}
The discovery of new particles and exotic states in high-energy physics often stems from the identification of rare or unexpected decays that deviate from known Standard Model processes. Particularly in diffractive processes and ultraperipheral collisions (UPCs) \cite{Baltz:2007kq}, such rare decays can manifest as strikingly clean signals, where a small number of final-state particles are produced with minimal background from additional partonic activity. These processes offer a unique opportunity to study both conventional hadrons and recently observed exotic hadrons - such as tetraquarks \cite{LHCb:2021vvq} and pentaquarks \cite{LHCb:2019kea} - in exclusive production modes. The detection of these exotic new hadrons has spurred renewed interest in studying rare hadronic states that challenge our understanding of quark dynamics within the framework of Quantum Chromodynamics (QCD).

Central exclusive production (CEP) in diffraction is characterised by the production of a small number of particles accompanied by rapidity gaps. Similarly, in UPC events, where the impact parameter between colliding ions is larger than the sum of their radii, photon-photon and photon-hadron interactions can lead to exclusive production of vector mesons and other particles. Machine learning can be used to perform online and offline selections of these events based on the topology of the events themselves \cite{Ragoni:2024jhg}. These events provide a particularly clean environment for detecting rare decays, as the underlying event activity is suppressed.

However, identifying such rare decays, particularly exotic hadrons, within the vast datasets of high-energy experiments like those at the Large Hadron Collider (LHC) and at the ePIC experiment at the future Electron-Ion Collider (EIC) remains challenging. Traditional methods, such as cut-based analyses, e.g. \cite{ALICE:2023jgu}, or supervised machine learning, e.g. \cite{ALICE:2022sco}, rely heavily on predefined decay signatures or labeled data. These approaches, while effective for known particles, are less suited for the discovery of new or unexpected processes, e.g. if the decay topology is rather complicated, as seen for pentaquark searches \cite{LHCb:2019kea}. Since diffractive and ultraperipheral events are very clean, with a small number of produced particles and low backgrounds, methods that can discern subtle differences between typical and anomalous decays are needed. 

In this work, we propose the use of an autoencoder neural network \cite{Hinton:2006tev} (a type of unsupervised learning model) to detect anomalies in exclusive production events. By training the autoencoder on typical decays, which are commonly produced in diffractive and ultraperipheral events, the network learns to reconstruct these events with low error. Rare decays are expected to deviate from the learned patterns, resulting in higher reconstruction errors, thus signaling potential anomalies.

The autoencoder technique is particularly well-suited for diffractive and ultraperipheral collision events due to the inherently clean signal environment, where the final state consists of only a few particles. By analyzing the reconstruction error and the corresponding invariant mass distributions, we demonstrate that this method can successfully separate typical decays from exotic hadronic states. This approach provides a new framework for anomaly detection in large datasets, with the potential to uncover new physics in the form of undiscovered particles or unexpected decay modes, particularly in exclusive production processes at the LHC and at the future EIC.

\section{Methodology}
In this section we describe a toy exercise carried out using synthetic data to illustrate the feasibility of the technique and clarify its application in real physics experiments at RHIC, the LHC and the future EIC.
\subsection{Generation of synthetic data}

To simulate a realistic dataset for anomaly detection in exclusive particle production, we employed a toy model that emulates both typical and rare decays observed in both CEP and UPC events. 

The typical decay processes simulated were as follows:
\begin{itemize}
    \item $\jpsi \rightarrow \mu^+\mu^-$: the momenta of the final state muons were generated assuming an isotropic decay in the center-of-mass frame;
    \item $\psip \rightarrow \mu^+\mu^-\pi^+\pi^-$: representing the decay mode $\psip \rightarrow \jpsi\pi^+\pi^-$, the momenta of the muons and pions were distributed uniformly, with no specific correlation between the decay products.
\end{itemize}
For the rare decay processes simulated, we focused on exotic hadrons that could be produced in exclusive events:
\begin{itemize}
    \item $\text{X(3872)} \rightarrow \mu^+\mu^-\pi^+\pi^-$: the momenta of the muons and pions were distributed uniformly, with no specific correlation between the decay products, as for the \psip;
    \item a structure with mass compatible with the X(6900): we generated two possible decay modes: (1) a four-muon final state and (2) a five-track final state. In the five-track mode, the tracks represented a combination of muons and pions, specifically three muons and two pions.
\end{itemize}

\begin{figure}[b]
\includegraphics[width=1.\columnwidth]{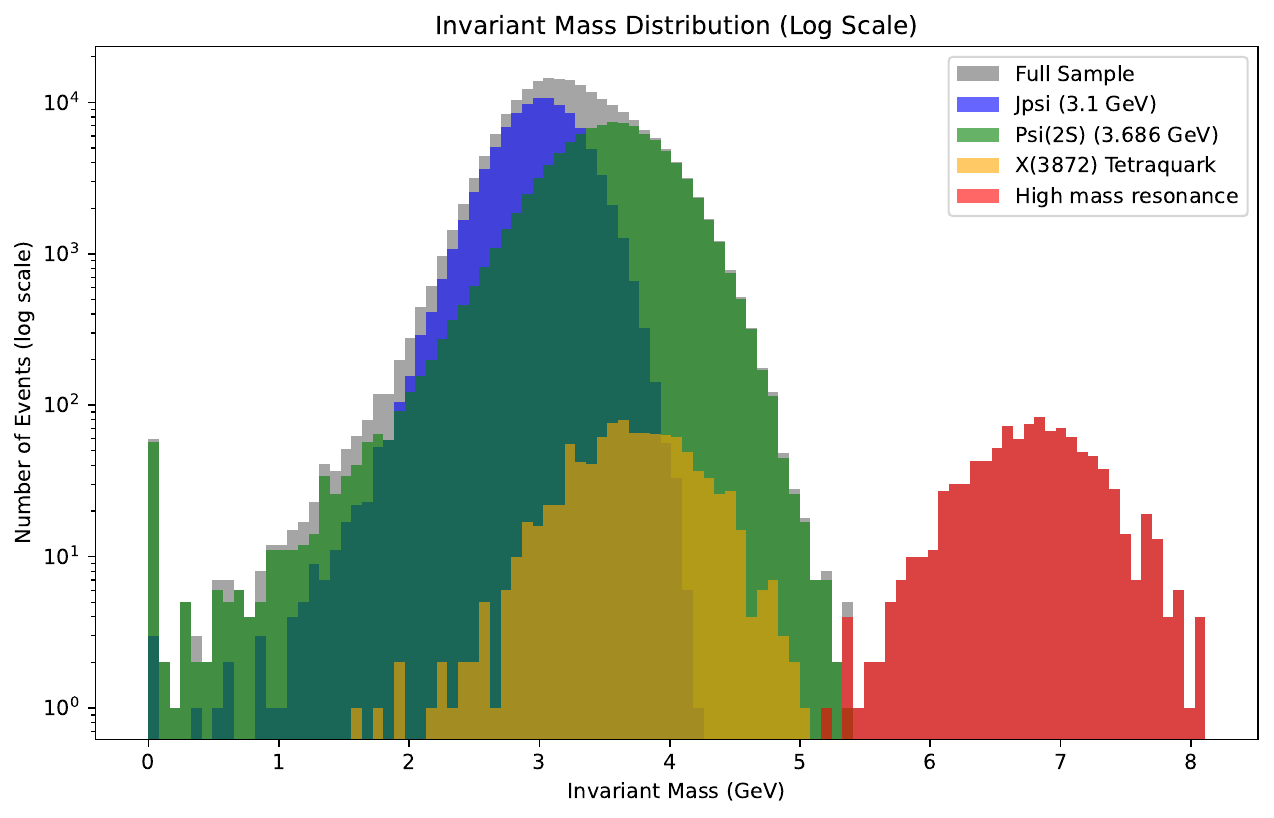}
\caption{\label{fig:cocktail} Invariant mass distribution of the full particle cocktail produced as synthetic data.}
\end{figure}

The full invariant mass distribution generated is shown in Fig.~\ref{fig:cocktail}, with different particle species represented in different colours. The target channels for the anomaly detection are generated as a sample that is 1\% of the size of the charmonia events. %
The features associated with each event were standardised to ensure consistent input for the machine learning model. Each event included the momenta \((p_x, p_y, p_z)\) and energy \(E\) of each particle, as well as the total invariant mass of the system. 

\subsection{Autoencoder architecture}

An autoencoder neural network \cite{Hinton:2006tev} was implemented to detect anomalous decays \cite{Farina:2018fyg}, using the \textit{Keras} library \cite{chollet2015keras}. An autoencoder is an unsupervised neural network model that learns to compress data into a lower-dimensional representation (latent space) and then reconstruct the original input from this compressed form. The network consists of two main components: an encoder, which reduces the dimensionality of the input data by extracting its most relevant features, and a decoder, which attempts to reconstruct the original data from this compressed representation. By training the model on typical events, such as \jpsi and \psip decays, the autoencoder learns to minimise the reconstruction error for these events. When applied to unseen data, i.e. events that differ significantly from the training data such as rare decays, or new particles, tend to have higher reconstruction errors, allowing for anomaly detection.

\begin{figure*}
\includegraphics[width=1\columnwidth]{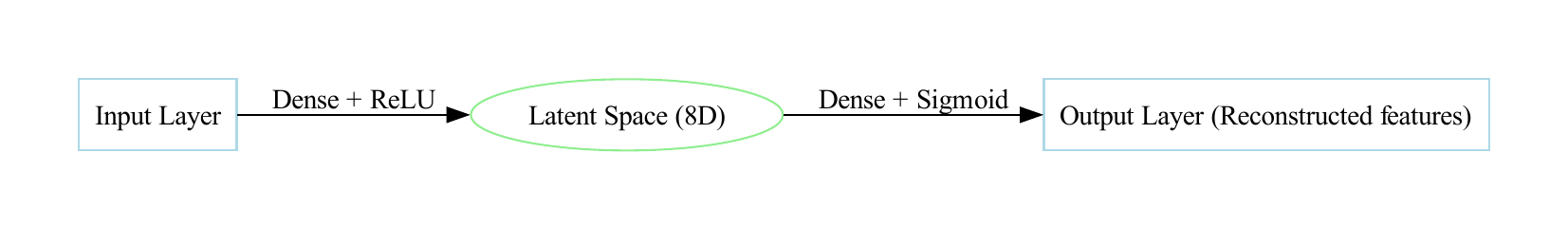}
\caption{\label{fig:autoencoder}Design of the autoencoder.}
\end{figure*}

The autoencoder was designed as follows:
\begin{itemize}
    \item \textbf{Input Layer}: The size of the input layer $\sigma$ is equal to $\sigma = 4 \times N + 1$, where $N$ is the maximum number of   daughter tracks among all resonances. In this case $N = 5$ from the high mass state, and we multiply $N$ by four to account for all the momentum components. When the number of daughter tracks is smaller than $N$, the feature vector is filled with zeroes for the missing tracks. There is one additional feature to account for the invariant mass of the event.
    \item \textbf{Encoder}: The encoder reduced the input to an 8-dimensional latent space using a fully connected Dense layer - every neuron of the layer is connected to the next layer - with a Rectified Linear Unit (ReLU) activation function, which is the standard choice for the encoder part of an autoencoder \cite{dertat}. The dimensionality of the latent space is a design choice to ensure that the encoder captures the key features of typical events.
    \item \textbf{Decoder}: The decoder reconstructed the input from the latent space using a Dense layer with sigmoid activation, producing a set of output features as close as possible to the original input.
\end{itemize}
The architecture of the autoencoder is shown in Fig.~\ref{fig:autoencoder}.%

The autoencoder was trained exclusively on the typical events to allow the model to learn their structure. 
The loss function used was the \textit{mean squared error (MSE)}, which calculates the squared differences between the input and reconstructed output:
\[
\text{MSE} = \frac{1}{\sigma} \sum_{i=1}^{\sigma} (x_i - \hat{x}_i)^2
\]
where $\sigma$ is the number of features defined above, $x$ is the input and $\hat{x}$ is the reconstructed feature vector. To prevent overfitting, which reduces the ability of the model to generalise to new data, early stopping is employed. This is carried out by halting the iterative procedure of the training if the performance of the model did not improve after 10 consecutive iterations, also known in literature as epochs.

\subsection{Anomaly detection via reconstruction error}
\begin{figure}[b]
\includegraphics[width=1.\columnwidth]{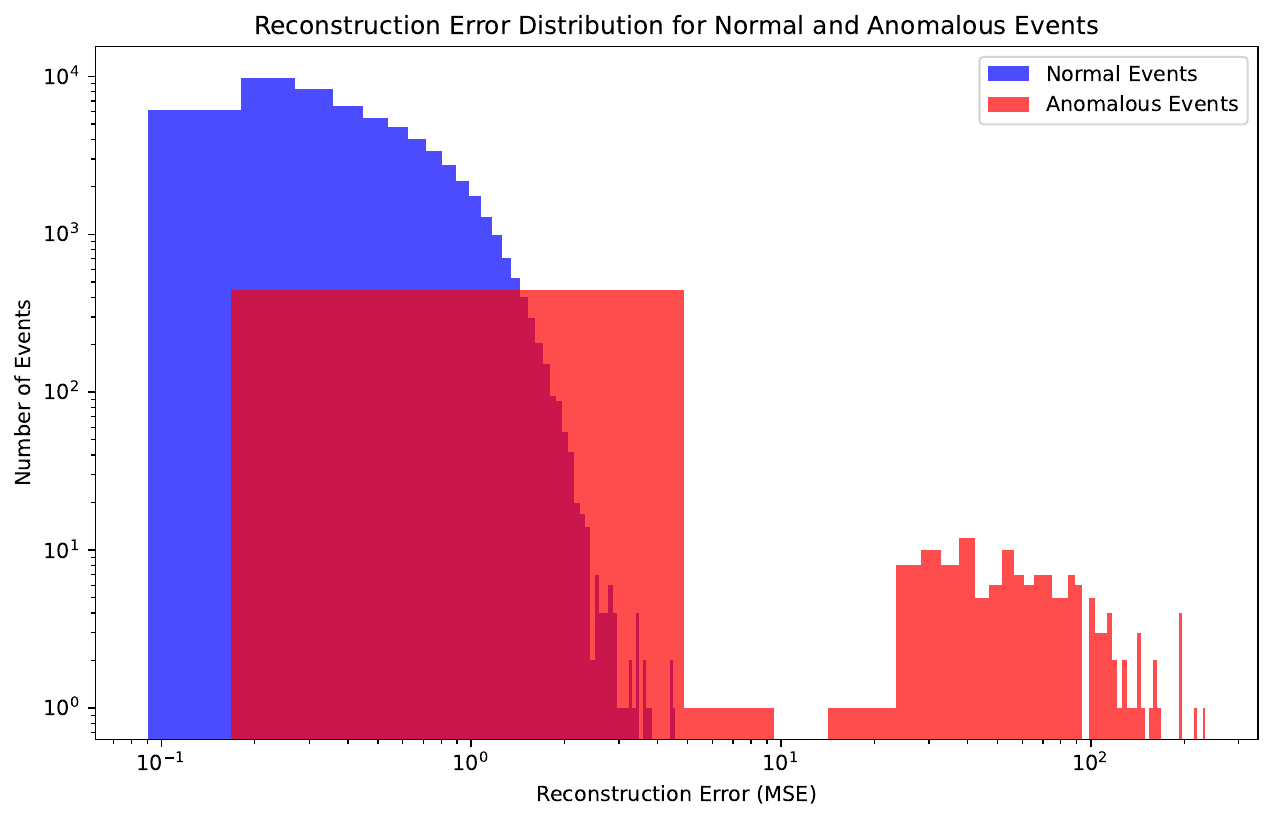}
\caption{\label{fig:mse} Distribution of the mean squared error (MSE) for the test events. Anomalous events can be flagged by the technique due to high values of MSE. The two categories of events are plotted with variable bin width to enhance the visibility of the low population of the anomalous events.}
\end{figure}

After training, the autoencoder was applied to both typical and rare decays in the test set. The reconstruction error, measured by the MSE, was used as the anomaly score. Figure~\ref{fig:mse} shows the distribution of the mean squared error (MSE) between the input and reconstructed test events. Owing to the design choice of eight dimensions for the latent space, the data are not overcompressed. This manifests with a low reconstruction error for typical events, with MSE values clustered around zero, indicating that the autoencoder successfully learned to reconstruct known decays like \jpsi and \psip with high accuracy. In contrast, rare decays, such as those corresponding to exotic hadrons, exhibit significantly higher MSE values, suggesting that these events deviate from the typical patterns learned by the model.

\begin{figure}[bh!]
\includegraphics[width=1.\columnwidth]{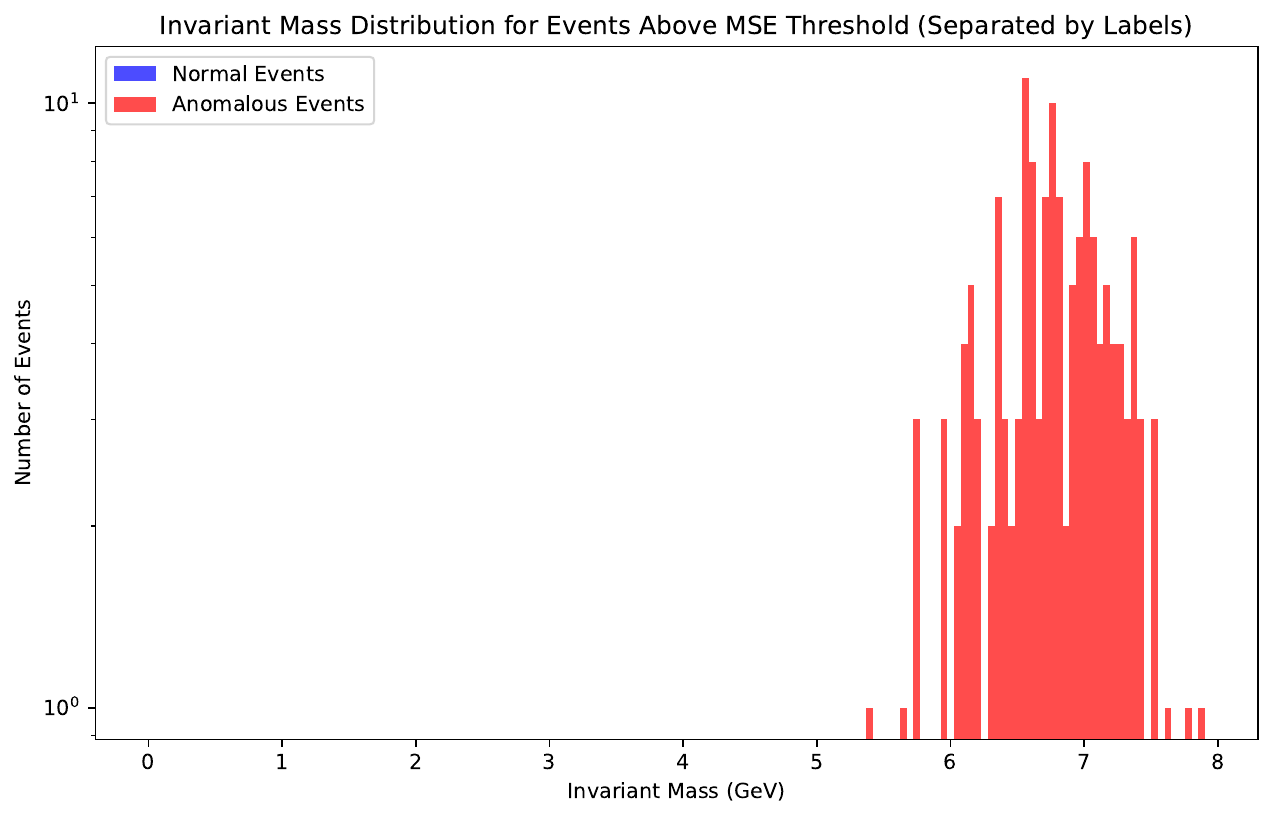}
\caption{\label{fig:flagged} Invariant mass distribution of those events which were flagged by the autoencoder as anomalous, i.e. with high MSE values.}
\end{figure}

Events with MSE above 10 were flagged as anomalous. Figure~\ref{fig:flagged} shows the invariant mass distribution of the events flagged as anomalous by the autoencoder. As expected, these events correspond predominantly to the injected rare decays. Notably, a peak near 6.9 GeV is visible. 

The autoencoder did not flag the X(3872) tetraquark in this test, as shown in the mass distribution of the flagged events. This indicates that the kinematics of the tetraquark are more similar to typical decays, making it harder for the autoencoder to distinguish this state based solely on reconstruction error. Further tuning of the network and exploration of additional features, such as angular correlations or momentum asymmetry, or even just particle identification capabilities, could improve the sensitivity to more exotic states. We foresee that this will be quite trivial in experiments since the autoencoder will be trained over the simulations convoluted with the detector geometry and its response. 

\section{Discussion}

The results presented in this work highlight the power and flexibility of autoencoders for anomaly detection in low-background environments such as diffractive and ultraperipheral collision (UPC) events. These environments, characterised by small numbers of final-state particles and suppressed hadronic activity, provide a unique opportunity to apply machine learning techniques for detecting rare processes or new particles, such as exotic hadrons.

One of the key advantages of using autoencoders is their ability to operate in an unsupervised fashion, which makes them well-suited for exploring new physics where no clear labels or predefined signals are available. Unlike traditional supervised learning methods that require labeled datasets, autoencoders learn the structure of the data from typical events and can automatically detect deviations that may signal rare or previously unknown processes. 

Anomalous events, characterised by high reconstruction error, indicate rare processes or signals which are not represented in the training set. This approach is particularly relevant for detecting rare processes in UPC and diffractive events, where the particle multiplicity is low. The autoencoder naturally handles such low-background environments by learning the normal kinematic features of typical processes and flagging significant deviations. Autoencoders also offer flexibility in their application. The dimensionality of the latent space can be adjusted to control the network's sensitivity, allowing for fine-tuning of the model for different types of physics events.

There are several advantages of using anomaly detection through autoencoders over more traditional searches using the detection of resonances in invariant mass distributions. Autoencoders make no assumptions about the decay channel itself, they have no intrinsic bias regarding which contributions are signal and which ones are background, and they enforce no cut on the available phase space. Hence, this approach reduces the risk of introducing biases since the autoencoder is not searching for a specific signal, but rather for deviations from normal behavior. Autoencoders learn the typical structure of the entire dataset, including background and known signals, and flag anomalies based on how different they are from this typical structure. This can lead to the discovery of signals in unexpected regions of the dataset that may have been overlooked with manual cut-based methods. Finally, while typical searches with invariant mass distributions involve looking at 1-dimensional, or at most 3-dimensional distributions, autoencoders operate at much higher dimensionalities. Autoencoders can then be considered one of the options to operate zero-bias new particle searches, especially in UPC and diffractive events.

\subsection{Application to CEP and UPC events}
The CEP and UPC event environments are particularly well-suited for the application of autoencoders because of the relatively low multiplicity of final-state particles and the clean event topology. In such events, subtle differences in particle kinematics, such as slight deviations in momentum or energy, can indicate the presence of new processes. In this exercise, showing an example using toy simulations, the autoencoder was trained on typical meson decays and then applied to test data that included both typical and rare exotic decays.  The autoencoder was able to flag these rare decays without prior knowledge of their presence, showing that this technique has the potential to enhance the discovery of new particles in low-background environments such as those characteristic of exclusive processes.
In this exercise the signature of the X(3872) decay was not flagged by the autoencoder. This highlights the need for a training on more features for the deployment in particle physics experiment, such as particle identification signals and angular distributions, to improve the sensitivity to these processes.
Finally, we would like to add that the use of autoencoders not only enabled anomaly detection in our dataset but also provided a significant reduction in data size. The autoencoder achieved a compression ratio of 2.5, corresponding to a 60\% reduction in disk space, while maintaining the essential features of the typical particle decays. This level of compression is particularly valuable when working with high-dimensional data like particle physics events, where each event contains multiple kinematic variables. Traditional lossless compression methods are often ineffective for such complex, high-dimensional floating-point data, making autoencoders a more suitable approach. The fact that the autoencoder was able to compress the data while still allowing accurate reconstruction demonstrates the model's ability to preserve crucial information. This balance between data reduction and information retention is critical for efficient storage, faster data processing, and effective anomaly detection, particularly in large-scale experiments like those at the LHC or future EIC, where (UPC) datasets can reach (several terabytes) petabytes in size. 
% \newline
% \newline

\subsection{Future Directions}
The results of this study show that autoencoders hold great promise for future applications in high-energy physics, particularly in the search for new physics beyond the Standard Model. Upcoming experiments at the Large Hadron Collider (LHC) and the future Electron-Ion Collider (EIC) will produce vast amounts of data from clean, exclusive processes. The use of autoencoders in this context can provide an effective and scalable zero-bias method for analyzing large datasets in real-time, flagging potential discoveries for further investigation. Furthermore, we note the potential for significant reduction of data storage needs which is achieved by the autoencoders: the key advantage of autoencoders is not solely saving space, but it is the ability to preserve the essential features of the data in a smaller representation.

In future work, exploring hybrid models that combine autoencoders with other unsupervised learning techniques or integrating physically motivated features into the autoencoder's input space could improve detection sensitivity.

\acknowledgments

This work was funded by the Department of Energy (DOE) of the United States of America (USA) through the grant DE-FG02-96ER40991.

% Bibliography

% [A] Recommended: using JHEP.bst file
\bibliographystyle{JHEP}
\bibliography{biblio.bib}

%% or
%% [B] Manual formatting (see below)
%% (i) We suggest to always provide author, title and journal data or doi:
%% in short all the informations that clearly identify a document.
%% (ii) please avoid comments such as "For a review'', "For some examples",
%% "and references therein" or move them in the text. In general, please leave only references in the bibliography and move all
%% accessory text in footnotes.
%% (iii) Also, please have only one work for each \bibitem.

% \begin{thebibliography}{99}

% \bibitem{a}
% Author,
% \emph{Title},
% \emph{J. Abbrev.} {\bf vol} (year) pg.

% \bibitem{b}
% Author,
% \emph{Title},
% arxiv:1234.5678.

% \bibitem{c}
% Author,
% \emph{Title},
% Publisher (year).

% \end{thebibliography}
\end{document}